\documentclass[conference,a4paper]{IEEEtran}
\usepackage{amsmath,epsfig, hyperref}
\usepackage[utf8]{inputenc}

\usepackage{ragged2e}
\usepackage[dvipsnames]{xcolor}
\usepackage{units}
\usepackage{boldline, soul}
\usepackage{multirow}
\usepackage{amsmath}
\usepackage{caption}
\usepackage{pgfplots}
\usepackage{tabularx}
\usepackage{multirow}
\usepackage{hhline}
\usepackage{balance}
\usepackage{booktabs}
\usepackage{balance}
\usepackage{soul}
\usepackage{amssymb}
\usepackage[space]{cite}
\usepackage{algorithm} 
\usepackage{algpseudocode} 
\usepackage{tikz}
\usepackage{wrapfig}
\usepackage{array}
\usepackage{dblfloatfix}

\usetikzlibrary{shapes.geometric, arrows}

\usepackage{makecell}
\setcellgapes{3pt}
\def\BibTeX{{\rm B\kern-.05em{\sc i\kern-.025em b}\kern-.08em
    T\kern-.1667em\lower.7ex\hbox{E}\kern-.125emX}}

\let\OLDthebibliography\thebibliography
\renewcommand\thebibliography[1]{
  \OLDthebibliography{#1}
  \setlength{\parskip}{0pt}
  \setlength{\itemsep}{0pt plus 0.3ex}
}

\usepackage{glossaries, soul}
\newacronym{rdo}{RDO}{rate-distortion optimization}
\newacronym{jvet}{JVET}{Joint Video Exploration Team}
\newacronym{vtm10}{VTM10}{VVC Test Model 10}
\newacronym{vtm15}{VTM15}{VVC Test Model 15}
\newacronym{vtm}{VTM}{VVC Test Model}
\newacronym{hevc}{HEVC}{High Efficiency Video Coding}
\newacronym{bdrate}{BD-rate}{Bjontegaard Delta-Rate}
\newacronym{qtmap}{QTdepthMap}{Quad Tree depth map}
\newacronym{cnn}{CNN}{Convolutional Neural Network}
\newacronym{ctc}{CTC}{Common Test Condition}
\newacronym{ragop32}{RAGOP32}{RandomAccess Group Of Picture 32}
\newacronym{ra}{RA}{RandomAccess}
\newacronym{vvc}{VVC}{Versatile Video Coding}
\newacronym{hdr}{HDR}{High Definition Range}
\newacronym{vr}{VR}{Virtual Reality}
\newacronym{qtmt}{QTMT}{quadtree with nested multi-type tree}
\newacronym{ctu}{CTU}{Coding Tree Unit}
\newacronym{ns}{NS}{No Split}
\newacronym{cu}{CU}{Coding Unit}
\newacronym{qt}{QT}{Quaternary Tree}
\newacronym{bt}{BT}{Binary Tree}
\newacronym{mt}{MT}{Multi-type Tree}
\newacronym{tt}{TT}{Ternary-type Tree}
\newacronym{hbt}{HBT}{Horizontal Binary Tree}
\newacronym{rdcost}{RDcost}{RateDistorsion Cost}
\newacronym{vbt}{VBT}{Vertical Binary Tree}
\newacronym{vtt}{VTT}{Vertical Ternary Tree}
\newacronym{htt}{HTT}{Horizontal Ternary Tree}
\newacronym{rf}{RF}{Random Forest}
\newacronym{mltcnn}{MLT-CNN}{multi-level tree CNN}

\newacronym{isp}{ISP}{Intra SubPartition}
\newacronym{satd}{SATD}{Sum of Absolute Transformed Difference}
\newacronym{jem}{JEM}{Joint Exploration Model}
\newacronym{bcnn}{B-CNN}{Branch Convolutional Neural Network}
\newacronym{mbmpcnn}{MBMP-CNN}{Multi-Branch Multi-Pooling CNN}
\newacronym{qp}{QP}{Quantization Parameter}
\newacronym{cbam}{CBAM}{Convolutional block attention module}
\newacronym{mae}{MAE}{Mean Absolute Error}

\newacronym{ts}{TS}{Time Saving}

\title{Light-weight CNN-based VVC Inter Partitioning Acceleration}
\iftrue
\author{
\IEEEauthorblockN{Yiqun Liu\IEEEauthorrefmark{1}\IEEEauthorrefmark{2}, Mohsen Abdoli\IEEEauthorrefmark{1}, Thomas Guionnet\IEEEauthorrefmark{1}, Christine Guillemot\IEEEauthorrefmark{2}, and Aline Roumy\IEEEauthorrefmark{2}
}

\IEEEauthorblockA
{
\IEEEauthorrefmark{1}Ateme, Rennes, France
}

\IEEEauthorblockA
{
\IEEEauthorrefmark{2}INRIA, Rennes, France
}
}
\fi

\begin{document}
\sloppy
\maketitle

\begin{abstract}
The \gls{vvc} standard has been finalized by \gls{jvet} in 2020. Compared to the \gls{hevc} standard, \gls{vvc} offers about 50\% compression efficiency gain, in terms of \gls{bdrate}, at the cost of about 10x more encoder complexity \cite{bross2021overview}. In this paper, we propose a \gls{cnn}-based method to speed up inter partitioning in \gls{vvc}. Our method operates at the \gls{ctu} level, by splitting each \gls{ctu} into a fixed grid of 8$\times$8 blocks. Then each cell in this grid is associated with information about the partitioning depth within that area. A lightweight network for predicting this grid is employed during the rate-distortion optimization to limit the \gls{qt}-split search and avoid partitions that are unlikely to be selected. Experiments show that the proposed method can achieve acceleration ranging from 17\% to 30\% in the \gls{ragop32} mode of \gls{vtm}10 with a reasonable efficiency drop ranging from 0.37\% to 1.18\% in terms of \gls{bdrate} increase.
\end{abstract}

\begin{figure*}[!t]
    \centering
    \includegraphics[width=0.8\linewidth]{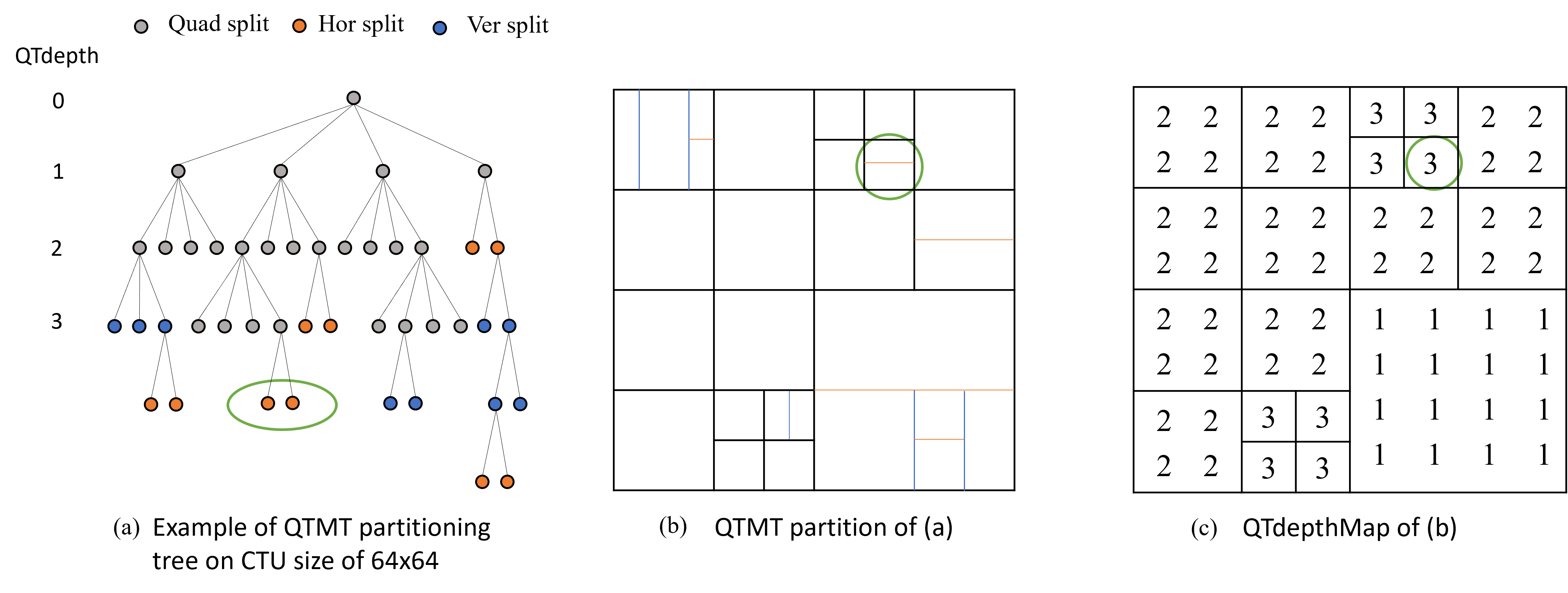}
    \caption{Example of partitioned 64x64 CTU with its QTdepthMap}
    \label{fig:fr_main}
\end{figure*}

\section{Introduction}

The design of efficient compression methods remains critical due to the continuously increasing demand for video transmission. In order to improve the efficiency of the coding, a new partition structure, called \gls{qtmt} \cite{huang2021block}, has been integrated into \gls{vvc}. In addition, multiple new inter coding modes \cite{wang2016adaptive} \cite{li2017efficient} \cite{alshin2010bi} have been adopted in \gls{vvc}. Thus, the inter prediction has been largely improved at the cost of increased complexity of \gls{rdo} of split types in inter coding. As a result, the newly adopted \gls{qtmt} and inter coding modes imply significant complexity for the partition search algorithm.

Various machine learning based methods \cite{amestoy2019tunable} \cite{kulupana2021fast} \cite{pan2021cnn} \cite{yeo2021cnn} have been proposed to speed up the \gls{rdo} search of inter partitioning in \gls{vvc}. They can be categorized into \gls{rf}-based and \gls{cnn}-based methods. In the first category, the work by Amestoy \textit{et al.} \cite{amestoy2019tunable} is a cascade of binary skip decisions predicted by trained \gls{rf}s. This method has been later improved by Kulupana \textit{et al.} \cite{kulupana2021fast}, by combining it with a hand-crafted early termination rule for the \gls{tt} splits. For \gls{cnn}-based methods, Pan \textit{et al.} propose in \cite{pan2021cnn} a multi-branch \gls{cnn}, namely MF-CNN, to perform a binary classification of the ``Partition" or ``Non-partition" at the \gls{cu} level, in order to early terminate the partitioning. In this paper, different models are trained for different \gls{cu} sizes. Furthermore, Yeo \textit{et al.} propose a method in \cite{yeo2021cnn}, where a variant of \gls{bcnn} \cite{zhu2017b}, called \gls{mltcnn}, is used to predict the partition mode of \gls{ctu}. The key feature of this \gls{cnn} is that its outputs correspond to different decisions at different levels of a split mode tree.

The aforementioned methods suffer from some limitations. \gls{rf}-based methods are complex in terms of number of trained models involved. For example, 18 models need to be trained separately for the different possible \gls{cu} sizes with obvious implication in terms of training and implementation complexity. The main disadvantage of \gls{cnn} methods is that their trade-offs between encoding complexity reduction and coding loss are much worse than those of \gls{rf} methods.

To address these issues, we develop a lightweight \gls{cnn}-based inter partitioning acceleration method for \gls{vvc} in this paper. Given that the partitioning syntax of \gls{vvc} can be specified as a series of QT-then-MT splits, the proposed method predicts the QTdepth of each 8$\times$8 block. The prediction is performed by a \gls{cnn} learned from a large dataset of encoded videos and is integrated into the \gls{vtm} with the help of a threshold to control the trade-off between complexity reduction and coding performance. To achieve accurate QTdepth prediction, we propose to base the decision on not only luminance values but also motion fields, and motion compensated residues. 

Compared to \gls{rf} methods, our proposed method achieves a similar performance. The main advantage is that our method is based on a unique model trained on \gls{ctu} level, which significantly eases the implementation of the method. Moreover, our lightweight method is scalable since we can adjust a threshold to obtain different levels of acceleration. Finally, the performance of the proposed method is significantly better than that of the related \gls{cnn}-based approaches. Furthermore, experiments show that our solution can be applied to different versions of the VTM with steady performance.

The remainder of this paper is organized as follows. Section II describes the proposed method in detail, while Section III presents results, and finally Section IV concludes the paper.

\section{Proposed method}

\subsection{QTdepthMap Representation}

Starting from the \gls{ctu} structure, \gls{vvc} performs \gls{rdo} for all possible split options as well as the \gls{ns} option, at each level of the partitioning tree, in order to find the partition which best exploits spatial and temporal redundancy. The \gls{qtmt} partitioning scheme in \gls{vvc} consists in splitting the \gls{cu} using either a \gls{qt} split or a \gls{mt} split. In the mode \gls{mt}, \gls{vvc} uses four split types: \gls{hbt} split, \gls{vbt} split, \gls{htt} split, \gls{vtt} split.

Fig.~\ref{fig:fr_main} (a) shows an example of a \gls{qtmt} partition tree, with its corresponding partition shown in Fig.~\ref{fig:fr_main} (b). Without loss of generality and for the sake of simplicity, the example in this figure is shown for a \gls{ctu} of size 64$\times$64 and \gls{qt}depth values in $\{0,1,2,3\}$, while in \gls{ctc} of \gls{vvc}, the \gls{ctu} of size 128$\times$128 results in \gls{qt}depth values in $\{0,1,2,3,4\}$. In Fig. \ref{fig:fr_main} (a), the red horizontal arc-shaped lines indicate the border between the last \gls{qt} split and the first \gls{mt} split (if any). 
According to the QT-then-MT scheme adopted in \gls{vvc}, once the first \gls{mt} split is chosen, the \gls{qt} split becomes forbidden for the following nodes. Therefore, one can represent partitioning series for reaching any arbitrary \gls{cu}, as N consecutive \gls{qt} splits, followed by M consecutive \gls{mt} splits. In the proposed method, we use the values of N that are associated with the leaf nodes (\textit{i.e.} final partitions) and form a grid called \gls{qtmap}. Fig. \ref{fig:fr_main} (c) shows  this map computed for the above partitioning example. The purpose of the paper is to propose a CNN architecture that predicts this \gls{qtmap}.

\subsection{Network architecture for \gls{qt}depthMap prediction}

\begin{figure*}
    \centering
    \includegraphics[width=0.9\linewidth]{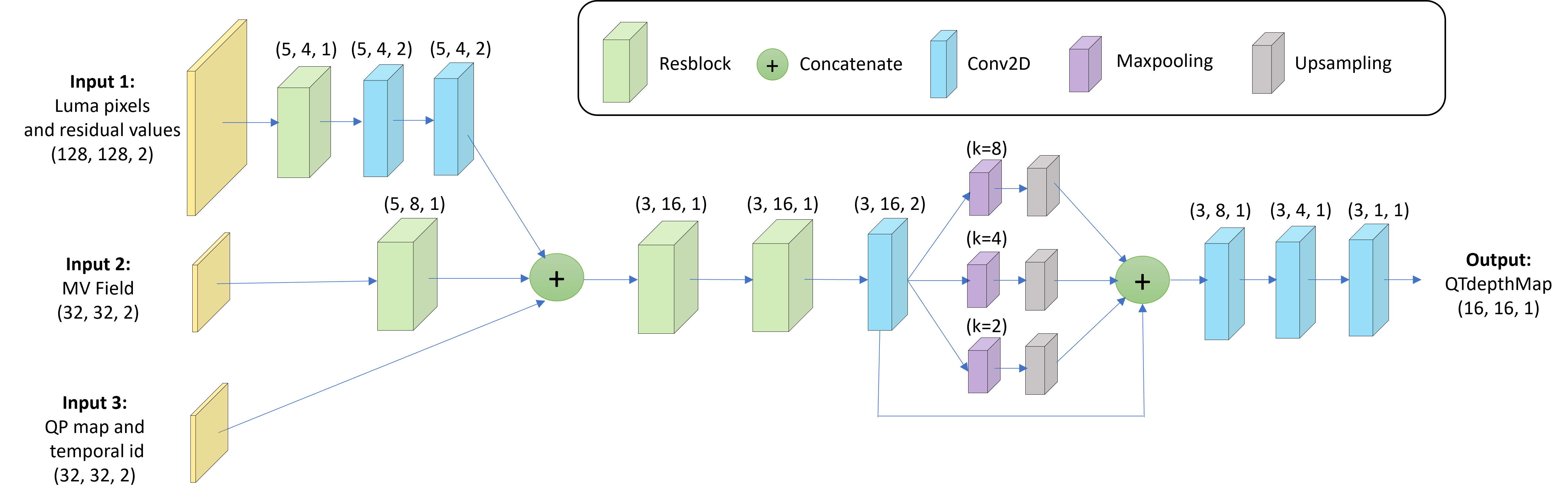}
    \caption{Structure of the proposed MBMP-CNN. The vector of three elements on top of Resblocks and Conv2D layers represents kernel size, number of filters and stride, respectively.  Value "k" denotes pooling size for Maxpooling layers.}
    \label{fig:mbmpcnn}
\end{figure*}

The \gls{cnn} proposed in this paper, called \gls{mbmpcnn}, is shown in Fig.~\ref{fig:mbmpcnn}. This network structure merges inputs of different types. For the first input, the luma values of \gls{ctu} are concatenated with residual blocks obtained by the motion-compensated prediction of the current \gls{ctu} from the nearest frame. The second input is formed by the motion vectors calculated on each 4x4 subblock in reference to the closest frame. The reason for choosing the residual block and the motion field as input is that these features reflect temporal correlation between frames as depicted in multiple papers \cite{amestoy2019tunable} \cite{pan2021cnn} \cite{yeo2021cnn} \cite{wang2018fast}. Finally, the \gls{qp} value and the temporal ID form the third input. With this third input, the proposed network achieves a good trade-off between prediction accuracy and memory consumption of the network. Indeed, a single learned network provides one prediction per QP value and per position of the frame in the GOP.

Tensors of same width and height corresponding to multiple input branches are concatenated before being fed into the main branch of the network architecture.The main branch begins with three residual blocks \cite{he2016deep} followed by the multi-pooling layer introduced in \cite{feng2021cnn}. The different kernel sizes are consistent with the \gls{cu} sizes at different \gls{qt}depths. Finally, three convolutional layers shrink the tensor dimension, which leads to the predicted 16$\times$16 \gls{qtmap} constituted by \gls{qt}depths on different 8$\times$8 sub-blocks. Each value equals to the \gls{qt}depth of the \gls{cu} containing the corresponding sub-block, as described earlier in Fig.~\ref{fig:fr_main} (c).

Inference within \gls{vtm} is performed by \gls{cnn} on each \gls{ctu} before entering the \gls{rdo} decision loop. By doing so, the encoder can use the acceleration algorithm presented in the next section to speed up the partition search. The CPU overhead of inference of this lightweight \gls{cnn} increases the encoding time by only 0.21\% when integrated into VTM10.

\subsection{Partitioning acceleration algorithm}

\begin{algorithm}
	\caption{MT and NS early skipping}
	 \textbf{Input:} \text{QTdepthMap;   QTdepth\textsubscript{cur}}; \text{  Th;   Size\textsubscript{CU};   Pos\textsubscript{CU}}\\
     \textbf{Output:} \text{SkipMT\_NS: Boolean to decide whether to skip  }\\\text{MT and NS split types or not}
	\begin{algorithmic}[1]
        \State Compute the QTdepth\textsubscript{avg} based on Size\textsubscript{CU}, Pos\textsubscript{CU} and QTdepthMap
        
        \If{$\text{QTdepth\textsubscript{avg}} > \text{(QTdepth\textsubscript{cur} + Th)}$}
        \State SkipMT\_NS = True
        \Else 
        \State SkipMT\_NS = False
		\EndIf
	\end{algorithmic} 
\end{algorithm} 

%We have merged the proposed \gls{qtmap} prediction with the original partition search process, with the goal of skipping the \gls{rdo} of unnecessary \gls{mt} and \gls{ns} splits. The algorithm for early skipping is depicted in Algorithm 1 and the overall algorithm is described by the flowchart in Fig.~\ref{fig:flowchart}.

\begin{figure}[!h]
    \centering
    \includegraphics[width=0.6\linewidth]{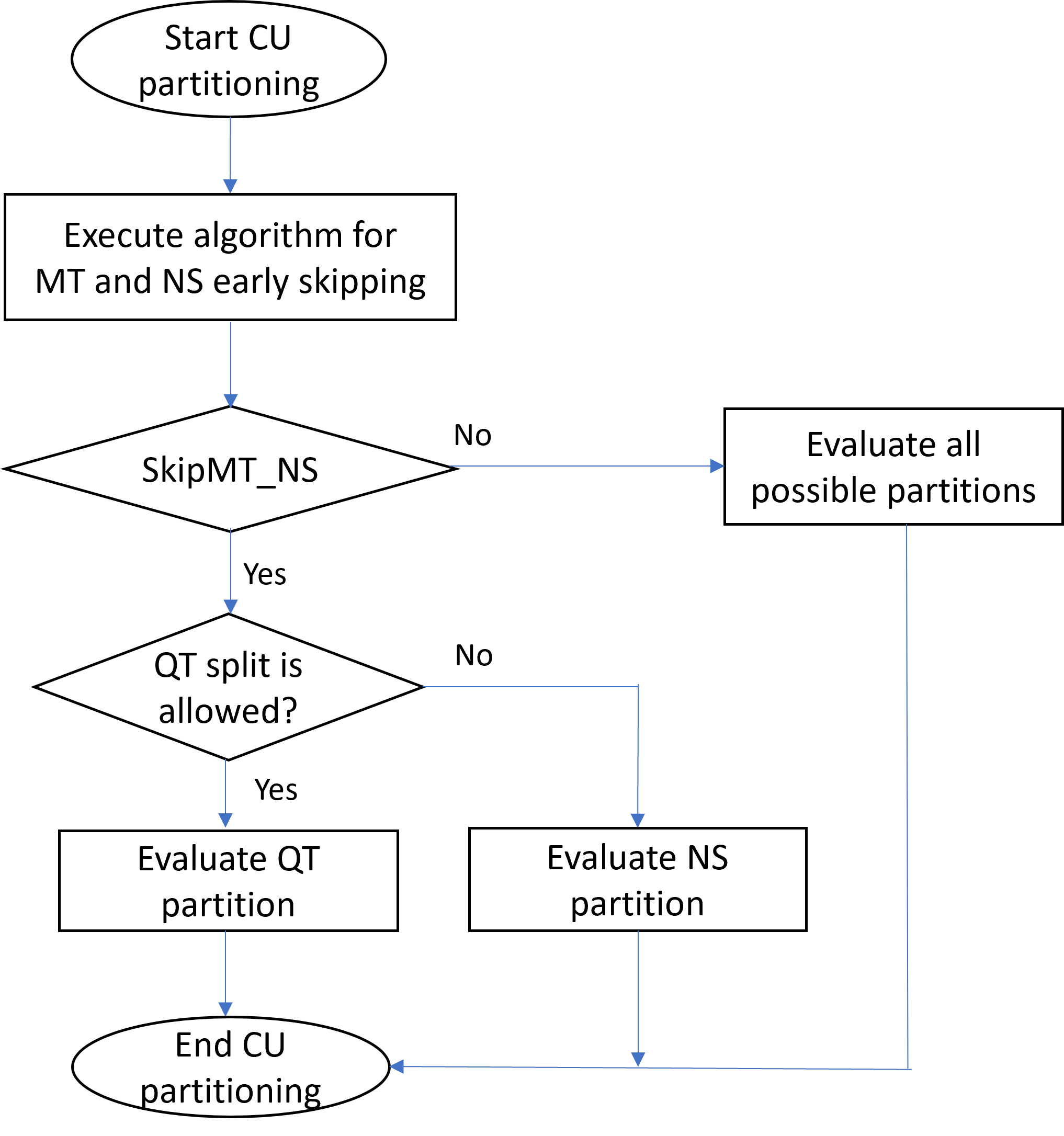}
    \caption{Flow chart of proposed acceleration method} 
    \label{fig:flowchart} 
\end{figure}

The main idea of the proposed acceleration algorithm is to avoid unnecessary \gls{mt} and \gls{ns} split checks in the QT-phase of the partitioning. Our algorithm navigates through the tree depicted in Fig.~\ref{fig:fr_main} (a) from top to bottom. Then, for each possible CU of the tree, if \gls{qtmap} predicts that a QT split must be performed, then we remove from the \gls{rdo} check all the \gls{mt} and \gls{ns} split checks. 

\begin{figure}[!h]
    \centering
    \includegraphics[width=0.9\linewidth]{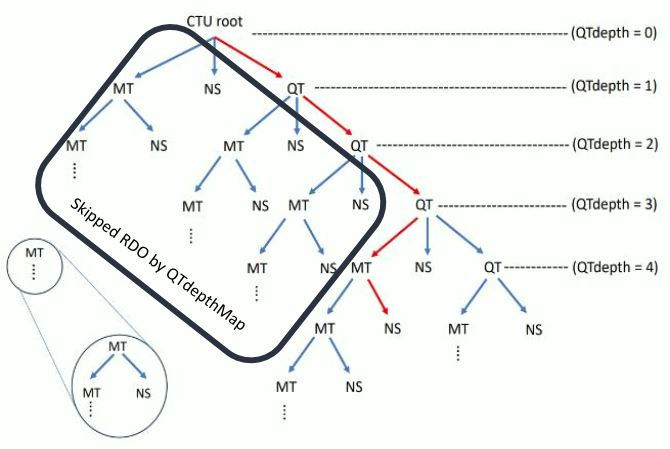}
    \caption{Example of acceleration by QTdepthMap} 
    \label{fig:illustration} 
\end{figure}

More precisely, consider a CU at depth QTdepth\textsubscript{cur}. Our predicted \gls{qtmap} provides a QTdepth value for all 8x8 subblocks of the \gls{ctu}. Therefore, for the CU under consideration, we compute QTdepth\textsubscript{avg}, the average of the QTdepth values given by the predicted \gls{qtmap} of all 8x8 subblocks inside the CU. Then, the predicted QTdepth of the CU is given by this QTdepth\textsubscript{avg} - Threshold. Finally, if the predicted QTdepth value (\emph{i.e.} QTdepth\textsubscript{avg} - Threshold) is greater than the QTdepth value of the CU (QTdepth\textsubscript{cur}), then a QT split is performed and all \gls{mt} and \gls{ns} split checks are avoided. We illustrate an example of acceleration through the prediction of QTdepthMap in Fig.~\ref{fig:illustration}. The circled \gls{cu} in Fig.~\ref{fig:fr_main} can be achieved by employing the partitioning outlined by the red path in Fig.~\ref{fig:illustration}. With the prediction of QTdepthMap, the \gls{qt}depth of the \gls{cu} can be determined as 3 before starting the partitioning process. Consequently, the \gls{rdo} check of partitions in the black box can be skipped.

The threshold value introduced above and denoted Th has two advantages. First, it allows us to compensate for outliers in the predicted QTdepthMap. Second, it allows to produce a scalable tradeoff between compression performance and acceleration. Indeed, a high threshold value leads to fewer skips in the RDO checks. Therefore, compression performance will be less affected, but acceleration will be reduced. The opposite effect occurs for small Threshold value. Note that such scalability cannot be achieved with methods based on the minimum and maximum values of \cite{feng2021cnn}.
The proposed method is summarized in Fig.~\ref{fig:flowchart}, where the early skipping strategy is detailed in Algorithm 1.

\subsection{Dataset and Training Details}

\begin{table*}[t]
\centering
\caption{\label{tab:Result} BD-rate and Time Saving tradeoff for the proposed method and state-of-the-art CNN-based methods. (The average of TS is geometric mean and that of BD-rate is arithmetic mean. Unit: \%)}        
\begin{tabular}{c|cc|cc|cc|cc}
\hline 
\hline
\multirow{2}{*}{Average on Class} & \multicolumn{2}{c|}{Pan \cite{pan2021cnn} \scriptsize{(VTM6)}} & \multicolumn{2}{c|}{Yeo \cite{yeo2021cnn} \scriptsize{(VTM11)}} & \multicolumn{2}{c|}{Proposed \scriptsize{(Th=0, VTM6)}} & \multicolumn{2}{c}{Proposed \scriptsize{(Th=0.1, VTM11)}} \\
                    & BD-rate            & TS            & BD-rate              & TS              & BD-rate            & TS            & BD-rate               & TS   \\
\hline 
\parbox[t]{2mm}{A} 	
																		    & 3.85 & 36.46 	& 1.95 & 16.99 	& 2.12 & 35.88 	& 1.47 & 26.05 \\
\hline 	                                                                                                                                                                                    
\parbox[t]{2mm}{B}  & 3.76 & 30.27 & 0.95 & 16.19 	& 1.59 & 31.79 & 1.06 & 23.34 \\
\hline                                                                                                                                                                                                      
\parbox[t]{2mm}{C} & 2.56 & 25.77 & 0.09 & 4.05 & 0.39 & 22.94 & 0.29 & 17.29 \\
\hline                                                                                                                                                                                                      
\parbox[t]{2mm}{D} & 2.35 & 21.53 	& 0.11 & 2.34 	& 0.19 & 12.34 	& 0.15 & 8.82 \\
\hline                                                                                                                                                                                                      
 \parbox[t]{2mm}{E}  & 2.81 & 35.15 &	NA	& NA		& 1.15 & 29.18 	& 0.80 & 14.23 \\
\hline                                                                                                                                                                                                      
																						\textbf{Total average} & \textbf{3.18} & \textbf{30.63} 	& \textbf{0.71} & \textbf{7.12} 	& \textbf{1.20} & \textbf{27.85} 	& \textbf{0.83} & \textbf{19.34} \\
 \hline
 \hline
\end{tabular}
\label{tab:result}
\end{table*}

The proposed method has been implemented in various VTM versions using the Frugally deep library \cite{frugallydeep2018} to load the trained model and to perform the inference on the CPU. 800 sequences of different resolutions, namely 240p, 540p, 720p and 1080p, have been encoded using the \gls{vtm10} with the \gls{ragop32} configuration. From this set, 600 sequences of 64 frames come from the BVI-DVC database \cite{ma2021bvi}, while the remaining sequences have been selected from the YouTube UGC dataset \cite{wang2019youtube}. We have collected the \gls{ctu}-level features, mentioned in previous section, together with its \gls{qtmap}. 200k \gls{ctu} samples have been randomly chosen for each resolution.
The model has been trained using Keras %\cite{chollet2015keras} 
on GTX1080ti with AMD 3700x processor. We use the L1 loss function and the Adam optimizer \cite{kingma2014adam}. The learning rate has been set to 1e$^{-3}$ for the first 20 epochs, then decreased by 10\% every 10 epochs. The batch size for training is 200. The model converges after 30 epochs.
%The \gls{mae} is used as validation metric.

\section{Experimental results}
The tests are carried out in \gls{ctc} sequences with the \gls{ra} configuration. \gls{bdrate} \cite{bjontegaard2001calculation} and \gls{ts} have been used to assess the performance. The formula to calculate \gls{ts} is given in Equation \ref{eq:ts}. T\textsubscript{Test} indicates the encoding time of the proposed method and T\textsubscript{VTM} indicates the encoding time of the encoder under the same condition.
\begin{equation}
TS = \frac{1}{4} \sum_{QP \in \{22,27,32,37 \}} \frac{T_{VTM}(QP) - T\textsubscript{Test}(QP)}{T_{VTM}(QP)}
\label{eq:ts}
\end{equation}

First, when test is conducted with VTM10, we obtain encoding time acceleration ranging from 17\% to 30\% for only 0.37\% to 1.18\% \gls{bdrate} loss in \gls{ragop32}. Second, since performances of different \gls{vtm}s vary significantly \cite{vtmcomplexity}, and in order to perform fair comparison with state of the art methods, the tests have been also conducted for other \gls{vtm} versions. Note, however, that our \gls{cnn} is trained on a dataset generated from \gls{vtm}10, which works against our method.

In Table \ref{tab:result}, the proposed method is compared with the two \gls{cnn}-based inter partitioning methods in \cite{pan2021cnn} and \cite{yeo2021cnn}. For a fair comparison,  we have implemented our method in \gls{vtm}6 and \gls{vtm}11. %Two values of 0 and 0.1 are used as the internal thresholds of the proposed method. 
The proposed method offers a much better trade-off between time saving and performance drop than the state-of-the-art methods. More precisely, when we set the threshold Th in Algorithm 1 to 0, we obtain a similar acceleration as the method developed in [8] with a significantly lower loss. Similarly, when the threshold Th equals 0.1, acceleration is tripled and the loss remains the same as in [9].

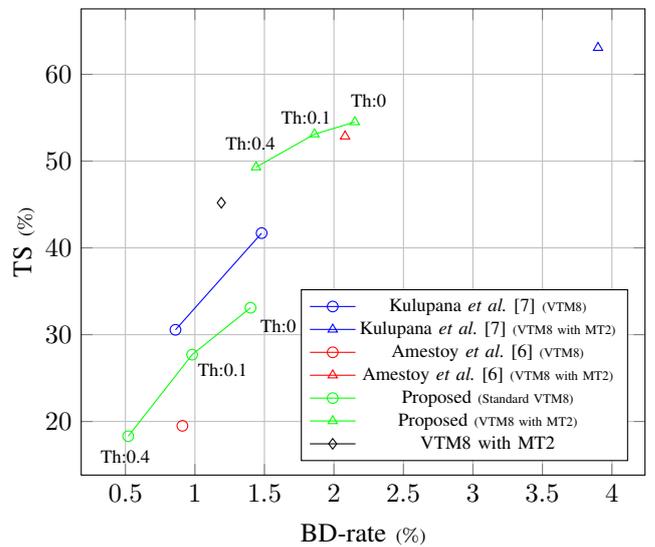
\begin{figure}[h]
    \centering
    \begin{tikzpicture}
\begin{axis}[
	xlabel={BD-rate \scriptsize{($\%$)}},
	ylabel={TS \scriptsize{($\%$)}},
	legend pos=south east,
	legend style={nodes={scale=0.7, transform shape}},
	width=9cm,
	y label style={at={(axis description cs:0.1,.5)},anchor=south},
	grid=major
]
\addplot[blue,mark=o] coordinates {
(0.86,	30.56)
(1.48,	41.7)
};

\addplot[blue,mark=triangle] coordinates {
(3.9,	63.06)
};

\addplot [red,mark=o] coordinates {
(0.91,	19.5)

};

\addplot [red,mark=triangle] coordinates {

(2.08,	52.86)
};

% \addplot coordinates {
% (3.18,	31)
% };

% \addplot[mark=square] coordinates {
% (0.7,	7.12)
% };

% Obsolete and wrong results

% \addplot[green,mark=*] coordinates {
% (0.42,	17)
% (0.85,	24)
% (1.04,	28)
% (1.21,  35)
% };

\addplot[green,mark=o] coordinates {

% th 0, 0.1, 0.4

(1.4, 33.1)
(0.98, 27.7)
(0.52, 18.3)

};

\addplot[green, mark=triangle] coordinates{

(1.44, 49.3)
(1.86, 53.1)
(2.15, 54.5)

};

% vtm standard with mt2
\addplot [black, mark=diamond] coordinates {
(1.19, 45.2)
};

\node [black] at (axis cs:0.5, 16) {\scriptsize{Th:0.4}};
\node [black] at (axis cs:1.2,	26) {\scriptsize{Th:0.1}};
\node [black] at (axis cs:1.6,  31) {\scriptsize{Th:0}};

\node [black] at (axis cs:2.25,  57) {\scriptsize{Th:0}};
\node [black] at (axis cs:1.8,  55) {\scriptsize{Th:0.1}};
\node [black] at (axis cs:1.4,  52) {\scriptsize{Th:0.4}};

% \node [red] at (axis cs:2.1,  50.5) {\scriptsize{MT2}};

% \node [blue] at (axis cs:3.7,  65) {\scriptsize{MT2}};

\legend{Kulupana \textit{et al.} \cite{kulupana2021fast} \scriptsize{(VTM8)}, Kulupana \textit{et al.} \cite{kulupana2021fast} \scriptsize{(VTM8 with MT2)}, Amestoy \textit{et al.} \cite{amestoy2019tunable} \scriptsize{(VTM8)}, Amestoy \textit{et al.} \cite{amestoy2019tunable} \scriptsize{(VTM8 with MT2)}, Proposed \scriptsize{(Standard VTM8)}, Proposed \scriptsize{(VTM8 with MT2)}, VTM8 with MT2}
\end{axis}
\end{tikzpicture}
    \caption{Complexity gains versus BD rate loss in comparison with \gls{rf} based methods.
    % \hl{Y-axis is complexity reduction. Harmonize it with the notation (i.e. TS)}
    }
    \label{fig:result_curve}
\end{figure}

The two other acceleration methods for inter partitioning of \gls{vvc}, notably the \gls{rf}-based works presented in \cite{amestoy2019tunable} and \cite{kulupana2021fast}, generate their results differently. The test set used in these papers is different from \gls{jvet} \gls{ctc}. Also, the encoder configuration is modified. In fact, the maximum of the \gls{mt} split depth has been reduced from its default value 3 to 2 in \cite{amestoy2019tunable} and \cite{kulupana2021fast}. We refer to this setup as VTM8 with MT2 in the following. Fig.~\ref{fig:result_curve} presents a performance comparison of \gls{rf} solutions with our method implemented in \gls{vtm}8 and VTM8 with MT2. For a fair comparison, the results in this figure are calculated for all sequences \gls{ctc} included in the test set of \cite{kulupana2021fast}. The label ``Th" indicate the value of the threshold. 

Interestingly, the proposed method achieves similar results to these of the \gls{rf} based methods. This is remarkable as our method requires a unique network to predict the QTdepthMap for any QP value and any frame position in the GOP, and the learning has been performed for a different VTM version from the one used for testing.

\section{Ablation Study}

\begin{figure}[h]
    \centering
    \begin{tikzpicture}
\begin{axis}[
	xlabel={BD-rate \scriptsize{($\%$)}},
	ylabel={TS \scriptsize{($\%$)}},
	legend pos=south east,
	legend style={nodes={scale=0.7, transform shape}},
	width=9cm,
	y label style={at={(axis description cs:0.1,.5)},anchor=south},
	grid=major
]

\addplot[black,mark=diamond] coordinates {
(0.6,	22)
(0.74,	25)
};

\addplot[blue,mark=asterisk] coordinates {
(0.75,	24)
};

\addplot [red,mark=o] coordinates {
(0.8,	25)

};

\legend{
MBMP-CNN, CNN\_No\_MVField, CNN\_No\_TempID
}
\end{axis}
\end{tikzpicture}
    \caption{Complexity gains versus BD rate loss for CNN models in ablation study}
    \label{fig:ablation}
\end{figure}
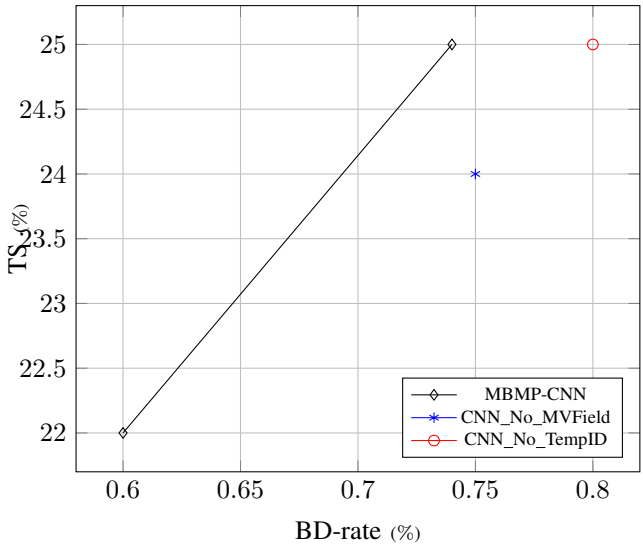

The \gls{ctu} pixel value and \gls{qp} value are essential for the \gls{cnn} model. In this work, we include additional input features such as residual value, tempID value, and MV Field. To assess the impact of these three features, we conducted an ablation study. Three \gls{cnn} models, named \emph{CNN\_No\_Resi}, \emph{CNN\_No\_MVField} and \emph{CNN\_No\_TempID}, were trained separately on the same dataset. Compared to our purposed model, \emph{CNN\_No\_Resi} shares the same CNN structure except that the residual input feature is removed, and a similar approach is taken for \emph{CNN\_No\_MVField} and \emph{CNN\_No\_TempID}. We then evaluated these models on \gls{ctc} sequences. The performances of these three models are compared to our \emph{MBMP-CNN} in Fig.~\ref{fig:ablation}. The performance of \emph{CNN\_No\_Resi} is 61\% TS with a 25.25\% BD-rate increase, indicating that the model malfunctions without the residual value as an input feature. Fig.~\ref{fig:ablation} shows that the performance of the CNN model is worse when removing MV Field and tempID features. Consequently, we can conclude that the residual value is crucial for predicting the QTdepth, and MV Field and tempID are valuable features for the CNN model.

\section{Conclusion}

In this paper, we have proposed a lightweight \gls{cnn}-based CU partition acceleration algorithm specialized for inter coding in \gls{vvc}. The proposed method adopts \gls{cnn}-based skip rule to effectively speed up the \gls{rdo} of \gls{mt} and \gls{ns} split types.
% The proposed \gls{mbmpcnn} takes inter related features as input and predicts a \gls{qtmap}. An early termination algorithm is proposed using the predicted \gls{qtmap}, to reduce the complexity of the \gls{mt} and \gls{ns} partitions. 
This scalable method succeeds in speeding up \gls{vtm10} efficiently from 17\% to 30\% for only 0.37\% to 1.18\% rate increase in the \gls{ragop32} configuration, which largely outperforms \gls{cnn} approaches. Compared with \gls{rf} based methods, we have attained the same level of performance with a method which is easier to implement.

% \bibliographystyle{alphanum} 

\bibliographystyle{IEEEbib}
\bibliography{icme2022final}

\end{document}